\documentclass[12pt]{article}

\usepackage{amsmath}
\usepackage{a4wide}
\usepackage{epsfig}
\usepackage{theorem}
\usepackage{amssymb}
\usepackage{bbm}
\usepackage[T1]{fontenc}
\usepackage[latin1]{inputenc}
\usepackage{tuemath}

\newcommand{\bi}{\bigskip}
\newcommand{\no}{\noindent}
\newcommand{\bea}{\begin{eqnarray}}
\newcommand{\eea}{\end{eqnarray}}
\newcommand{\hk}{\hspace{0.1cm}}

\newcommand{\lk}{\left(}

\newcommand{\sli}{\sum\limits}

\newcommand{\il}{\int\limits}

\renewcommand{\vec}[1]{\mbox{\boldmath$#1$\unboldmath}} 

\numberwithin{equation}{section}

\begin{document}

\title{On Tomonaga's Theory of Split-Anode Magnetrons}

\author{W. Dittrich\\
Institut f\"ur Theoretische Physik\\
Universit\"at T\"ubingen\\
Auf der Morgenstelle 14\\
D-72076 T\"ubingen\\
Germany\\
electronic address: qed.dittrich@uni-tuebingen.de
}
\date{\today}
%


\maketitle
\bi

\no

\begin{abstract}
This article is meant to formulate the equations of motion of an electron in a cavity magnetron using action-angle variables.
This means following the electron's path on its way from a cylindrical cathode moving toward a co-axial cylindrical anode in 
presence of a uniform magnetic field parallel to the common axis. After analyzing the situation without coupling to an external 
oscillatory electric field, we employ methods of canonical perturbation theory to find the resonance condition between the 
frequencies of the free theory $\omega_r, \omega_\varphi$
and the applied perturbing oscillatory frequency $\omega$. A long-time averaging process 
will then eliminate the periodic terms in the equation for the now time-dependent action-angle variables. The terms that
are no longer periodic will cause secular changes so that the canonical action-angle variables $(J, \delta)$ change in a way 
that the path of the electron will deform gradually so that it can reach the anode. How the ensemble of the initially 
randomly distributed electrons forms spokes and how their energy is conveyed to the cavity-field oscillation is the 
main focus of this article. Some remarks concerning the importance of results in QED and the invention of radar theory 
and application conclude the article.

\end{abstract}

\section{Introduction}

A magnetron is a device for generating microwave power in the $10^9 - 10^{11}$ Hz range. (The frequency of radiation used in microwave 
ovens is approximately $2.5 \times 10^9$ Hz, with a wavelength of 12 cm.) The invention of the magnetron greatly influenced the technology
of radio-based detection and ranging methods (RADAR) in Europe during  tragic WWII.
In particular, the invention of the cavity magnetron by John Randell and Harry Boot of Birmingham University in early 
1940 marked a major advance in radar capability. The cavity magnetron was perhaps the single most important invention in the
history of radar in the 20th century. In the Tizard Mission in September of 1940, the functioning of the cavity magnetron was 
revealed to the U.S. in exchange for magnetron production facilities. The transfer of the British magnetron secrets to the U.S. 
was later described as ''the most valuable cargo ever brought to the shores of the United States.'' Shortly thereafter, the 
Radiation Laboratory based at MIT was created to further develop the device and usage. Besides I. Rabi and others, Julian 
Schwinger was one of the members of MIT's Radiation Lab. Needless to say, he was actively involved in wave propagation with 
radar devices. But in 1947, using the repurposed microwave-frequency electronics left over from his wartime radar work, Willis 
Lamb of Columbia University in New York measured a tiny shift - of about one part in a million - in the energy levels of an 
electron in the $2s$ and $2p$ orbitals of a hydrogen atom. Lamb's remarkable achievement  challenged physicists' prevailing 
understanding of the quantum vacuum. Schwinger managed to calculate the effects of quantum fluctuations on the electron's 
energy levels and obtain an answer that matched Lamb's measurement to an extraordinary precision. As it turned out, Japanese 
physicist Sin-Itiro Tomonaga had accomplished the same goal a few years earlier. Tomonaga's work on radar during WWII had 
proven similarly essential to his theoretical approach. It is this point in Tomonoaga's life that is the main focus of the 
present paper. Not Tomonaga's contribution to quantum electrodynamics but his wartime work on the theory of the cavity 
magnetron is at the center of our review article.

\section{Introducing action-angle variables}\label{section2}

We want to introduce action-angle variables by way of studying the behavior of an electron in a homogeneous magnetic field $B$. We
can be brief here because an introduction to this problem is explicitly outlined in reference \cite{1} on pages 105-108.

So let us start with the Hamiltonian
\begin{align}
 \label{2.1}
 H &= \frac{1}{2m} \left[ p^2_r + \lk \frac{p_\varphi}{r} - \frac{e A_\varphi}{c} \right)^2 \right] \, , \quad \quad A_\varphi  = 
 \frac{Br}{2} \, , \nonumber\\
 &= \frac{1}{2m} \left[ p^2_r + \lk \frac{p_\varphi}{r} - \frac{e B r}{2c} \right)^2 \right]  \, .
\end{align}
The electron motion is given by $re^{i \varphi (t)}$, which we want to express in terms of action-angle variables $(w,J)$.

Realizing that $H$ and $p_\varphi$ are constants of motion, we can write for the Hamilton-Jacobi equation:
\begin{align}
 \label{2.2}
 \frac{1}{2m}  \left\{ \lk \frac{\partial W}{\partial r} \right)^2 + \left[ \frac{1}{r} \frac{\partial W}{\partial \varphi} - 
 \frac{m \omega_c r}{2} \right]^2 \right\} = E = \alpha_1 \, .
\end{align}
Here we introduced the cyclotron frequency  $\omega_c = eB/mc$.

For $e >0$ and $B >0$ the motion is clockwise.

The characteristic function $W$ can be separated according to.
\begin{align}
 \label{2.3}
 W = W (r, \varphi; \alpha_1, \alpha_2) = \varphi p_\varphi + W_r (r) \, , \quad \quad
 \alpha_1 = E \, , \quad \quad \alpha_2 = p_\varphi \, , 
 \end{align}
where $W_r (r)$ satisfies the differential equation
\begin{align}
 \frac{d W_r}{d r} = \frac{\partial W_r}{\partial r} = \left[ 2 m E - 
 \lk \frac{p_\varphi}{r} - \frac{m \omega_c r}{2} \right)^2 \right]^{1/2} \, . \nonumber
\end{align}
Therefore we have to calculate the $r$ integral in
\begin{align}
 \label{2.4}
 W = \varphi p_\varphi + \il^r d r \left[ 2 m E - \lk \frac{p_\varphi}{r} - \frac{m \omega_c r}{2} \right)^2 \right]^{1/2} \, .
\end{align}
Note that for one period the angle $\varphi$ changes in negative direction by an  
amount $2 \pi$, so in $J\varphi  =  \oint p_\varphi d \varphi$ the area enclosed is below the $\varphi$
axis but in negative direction,
so that $J_\varphi =  2 \pi | p_\varphi|$ with $p_\varphi  <0$. This remark becomes important when calculating
\begin{align}
 \label{2.5}
 J_r = \oint p_r d r = \oint \frac{\partial W}{\partial r} d r = \oint \sqrt{2 m E - \lk \frac{p_\varphi}{r} - \frac{m \omega_c r}{2} \right)^2
 } d r \, .
\end{align}
The details of the calculation can be looked up in ref. \cite{1}. We then find 
$(E = H = \alpha_1)$:
\begin{align}
 \label{2.6}
 J_r = \pi \left[ \frac{2 H}{\omega_c} + p_\varphi - |p_\varphi| \right] \, , \quad \quad(p_\varphi < 0) \, ,
 \end{align}
so that $H$ can be expressed in terms of the action variables $(J_r, J_\varphi)$ as
\begin{align}
 \label{2.7}
 H = \nu_c (J_r + J_\varphi) \, , \quad \quad \nu_c = \frac{\omega_c}{2 \pi} \, .
\end{align}
The corresponding frequencies are given by
\begin{align}
 \label{2.8}
 \nu_r = \frac{\partial H}{\partial J_r} = \nu_c \, , \quad \quad
 \nu_\varphi = \frac{\partial H}{\partial J_\varphi} = \nu_c \, , \quad \quad \mbox{i.e.} \, , \quad \quad
 \nu_r = \nu_c = \frac{\omega_c}{2 \pi} \, .
 \end{align}
So far we found using $\frac{m \omega_c}{\pi} = \frac{e B}{\pi c}$
\begin{align}
 \label{2.9}
 W = - \frac{J_\varphi}{2 \pi} \varphi + \il^r d r \left[ \frac{m \omega_c}{\pi}  (J_r + J_\varphi) - \lk \frac{J_\varphi}{2 \pi r} - 
 \frac{e Br}{2 c} \right)^2 \right]^{1/2} \, .
\end{align}
 From this expression we deduce the angle variable
\begin{align}
 \label{2.10}
 w_r &= \frac{\partial W_r}{\partial J_r} = \il^r d r \frac{\partial}{\partial r} \left[ \frac{eB}{\pi c}
 (J_r + J_\varphi) - \lk \frac{J_\varphi}{2 \pi r} - \frac{e Br}{2 c} \right)^2\right]^{1/2} \nonumber\\
 &= \frac{1}{2 \pi} \arcsin \lk \frac{\frac{\pi e B}{2 c} r^2 - (J_r + \frac{1}{2} J_\varphi)}{\sqrt{J_r (J_r + J_\varphi)}} \right)
 =: \frac{1}{2 \pi} \arcsin F
\end{align}
or $w_r = \frac{1}{2 \pi} \arcsin F = \nu_r t + \beta_r = \frac{\omega_c}{2 \pi} t + \beta_r$.

$\beta_r$ is a phase constant independent of time - so is $\beta_\varphi$.
 
\begin{figure}
 \centerline{
 \includegraphics[width=.6\textwidth]{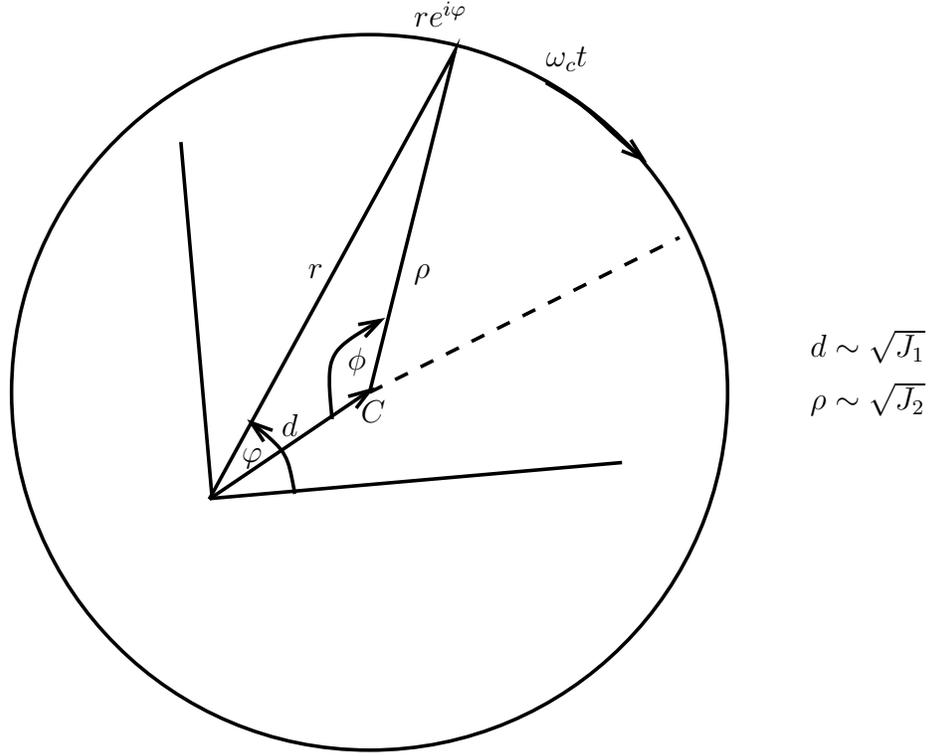}
 \put(30,150){$d \sim \sqrt{J_1}$}
  \put(30,130){$\rho \sim \sqrt{J_2}$}
 \put(-185,110){$\varphi$}
 \put(-170,120){$d$}
 \put(-120,180){$\rho$}
 \put(-160,180){$r$}
  \put(-140,125){$C$}
  \put(-145,145){$\phi$}
  \put(-120,275){$r e^{i \varphi}$}
    \put(-70,260){$\omega_c t$}}
  \caption{Circular motion of electron around center $C$.}\label{Fig.1}
\end{figure}

In equation (\ref{2.10}) we meet $r^2$. So let us solve for $r^2$:
\begin{align}
 \sin 2 \pi w_r &= F = \frac{\frac{\pi e B r^2}{2c} - (J_r + \frac{1}{2} J_\varphi)}{\sqrt{J_r (J_r + J_\varphi)}} \nonumber\\
 \mbox{or} \quad \frac{\pi e B r^2}{2 c} &= (J_r + \frac{1}{2} J_\varphi) + \sqrt{J_r (J_r + J_\varphi)} \sin 2 \pi w_r \, , \nonumber
\end{align}
from which we obtain the important result
\begin{align}
 \label{2.11}
 r^2 = \frac{2 c}{\pi e B} \left[ \lk J_r + \frac{1}{2} J_\varphi \right) + \sqrt{J_r (J_r + J_\varphi)} \sin 2 \pi w_r \right] \, .
\end{align}
 Noticing the following relation
\begin{align}
 \label{2.12}
 & 2 \left[ \lk J_r + \frac{1}{2} J_\varphi \right) + \sqrt{J_r (J_r + J_\varphi)} \sin 2 \pi w_r \right] \nonumber\\
 & \lk \sqrt{J_r + J_\varphi} - i \sqrt{J_r} e^{i 2 \pi w_r} \right) \lk \sqrt{J_r + J_\varphi} + i \sqrt{J_r} e^{- i 2 \pi w_r} \right) 
\end{align}
  we finally obtain the sought-after electron path in terms of action-angle variables:  
 \begin{align}
  \label{2.13}
  r e^{i \varphi} = \sqrt{\frac{c}{\pi e B}} \lk \sqrt{J_r - J_\varphi)} - i \sqrt{J_r} e^{i 2 \pi w_r} \right) \, .
 \end{align}
The physical meaning of this rather unfamiliar formula will become clearer when we introduce one more canonical 
transformation. But first let us complete our calculation by computing
\be
\label{2.14}
w_\varphi = \frac{\partial W}{\partial J_\varphi} = - \frac{\varphi}{2 \pi} + \il^r \frac{\partial}{\partial J_\varphi}
\left[ \frac{e B}{\pi c} (J_\varphi + J_\varphi) - \lk \frac{J_\varphi}{2 \pi r}  - \frac{e Br}{2 c} \right)^2 \right]^{1/2} dr \, .
\ee
With the definition
\be
\label{2.15}
G = \frac{\lk J_r + \frac{1}{2} J_\varphi \right) - \frac{c J^2_\varphi}{2 \pi e B r^2}}{\sqrt{J_r (J_r + J_\varphi)}}
\ee
we obtain
\begin{align}
 \label{2.16}
 w_\varphi &= \frac{- \varphi}{2 \pi} + \frac{1}{4 \pi} \arcsin F - \frac{1}{4 \pi} \arcsin G \nonumber\\
 &= \nu_\varphi + \beta_\varphi = \frac{\omega_c}{2 \pi} t + \beta_\varphi \, .
 \end{align}
Finally we end up with
\begin{align}
 \label{2.17}
 \varphi (t) = - 2 \pi w_\varphi + \pi w_r - \frac{1}{2} \arcsin \left[\frac{\lk J_r + \frac{1}{2} J_\varphi \right) \left\{
 J_r + \frac{1}{2} J_\varphi + \sqrt{J_r  (J_r + J_\varphi)} \sin 2 \pi w_r \right\} - \frac{J^2_\varphi}{4}}{\sqrt{ 
 J_r (J_r + J_\varphi)} \left\{ J_r + \frac{1}{2} J_\varphi + \sqrt{J_r (J_r + J_\varphi)} \sin 2 \pi w_r \right\}} \right] 
 \end{align}
 with
 \be
 w_\varphi = \frac{\omega_c}{2 \pi} t + \beta_\varphi \quad \mbox{and} \quad w_r = \frac{\omega_c}{2 \pi} t + \beta_r \, .
 \ee
In order to further simplify our  complicated formulae for a rather well-known problem, let 
us introduce the following canonical transformation
\be
\label{2.19}
J_1 = J_r + J_\varphi \, , \quad \quad J_2 = J_r
\ee
with the generating function of type $F_2 (q_1, q_2; P_1, P_2)$:
\be
\label{2.20}
F_2 (w_r, w_\varphi; J_1, J_2) = w_r J_2 + w_\varphi (J_1 - J_2)
\ee
so that we obtain the relation
\be
\label{2.21}
w_1 = \frac{\partial F_2}{\partial J_1} = w_\varphi \, , \quad \quad w_2 = \frac{\partial F_2}{\partial J_2} = 
w_r - w_\varphi \, : \quad \quad w_r = w_1 + w_2 \, , \quad \quad w_\varphi = w_1 \, .
\ee
Our former results then take the form
\begin{align}
 \label{2.22}
 H &= \frac{e B}{2 \pi m c} (J_r + J_\varphi) = \frac{\omega_c}{2 \pi} J_1 \, , \quad \quad \nu_1 = \frac{\partial H}{\partial J_1}
 = \frac{\omega_c}{2 \pi} \, , \quad \quad \nu_2 =  0 \, , \nonumber\\
 w_1 &= \nu_1  t + \beta_1 \, , \quad \quad w_2 = \beta_2 = const. \\
 \label{2.23}
 r^2 &= \frac{1}{\pi m \omega_c} \left[J_1 + J_2 + 2 \sqrt{J_1 + J_2} \sin 2 \pi (w_1 + w_2) \right] \, , \quad \quad\omega_c = \frac{e B}{m c} \, , \\
 \label{2.24}
 r e^{i \varphi} &= \sqrt{\frac{c}{\pi e B}} \lk \sqrt{J_1} - i \sqrt{J_2} e^{i 2 \pi (w_1 + w_2)} \right) 
\end{align}
with
\be
2 \pi  (w_1 + w_2) = 2 \pi \nu_1 t + const = \omega_c t + const. .
\ee
Formula (\ref{2.23}) allows for a simple graphical interpretation. For this reason we make use of the law of cosines in (Fig. \ref{Fig.1}):
\begin{align}
 \label{2.25}
 r^2 &= \rho^2 + d^2 - 2 \rho d \cos \phi \, , \quad \quad \rho^2 = \frac{J_2}{\pi m \omega_c} \, , \quad \quad
 d^2 = \frac{J_1}{\pi m \omega_c} \\
 \label{2.26}
 \cos \phi &= - \sin \lk \phi - \frac{\pi}{2} \right) \, , \quad \quad \phi - \frac{\pi}{2} = 2 \pi (w_1 + w_2) = \omega_c t + const.\nonumber\\
 \cos \lk \omega_c t + \frac{\pi}{2} \right) &= - \cos \omega_c t \nonumber\\
 \mbox{i.e.}, \quad \quad r^2 &= \rho^2 + d^2 + 2 \rho d \cos \omega_c t \, .
\end{align}
Also putting
\be
- 2 \pi w_\varphi + \pi w_r = - 2 \pi w_1 + \pi (w_1 + w_2) = \pi (w_2 - w_1)\nonumber
\ee
yields
\be
\label{2.27}
\varphi (t) = \pi (w_2 - w_1) - \frac{1}{2} \arcsin \left[ 
\frac{(J_1 + J_2) \sin 2 \pi (w_1 + w_2)  + 2 \sqrt{J_1 J_2}}{(J_1 + J_2) + 2 \sqrt{J_1 J_2} \sin 2 \pi (w_1 + w_2)} \right] 
\ee

\section{Tomonaga's theory of split-anode magnetron}

In this section we want to present the oscillation mechanism of a split-anode magnetron following the pioneering paper 
by Sin-itiro Tomonaga \cite{2}. The analytical method used follows very closely the procedure as given in the former section. 
Our aim is to clarify how electrons move in the inner space of the magnetron and, in particular, how the energy is conveyed 
from the electrons to the electric oscillation occuring in the resonant cavities of the split anode. 
Furthermore, we want to know how the phases of the circulating electrons which are originally distributed with equal 
probability over all values between zero and $2 \pi$ are forced into moving in phase, making several clusters of space 
charge in the magnetron and thus inducing electric oscillations in the split anode.

First of all we want to find the path of motion of an electron in a magnetron in the absence of an oscillatory 
electric field. Let $U_a$ be the anode voltage and $H$ the applied constant magnetic field, which we take to be parallel 
to the axis of the magnetron. Next, Tomonaga assumes for the potential energy of the electron at the position $(r, \varphi)$
\be
\label{3.1}
V (r, \varphi) = - e U_a \cdot \lk \frac{r}{r_a} \right)^2 \, ,
\ee
where $r$ is the distance from the center of the magnetron and $r_a$   denotes the anode radius. 
Making some minor changes in the notation used in the former section, we obtain for the Hamiltonian of the electron:
\be
\label{3.2}
H = \frac{1}{2 m} \left[ p^2_r + \lk \frac{p_\varphi}{r} + \frac{e H}{2 c} r \right)^2 \right] + V (r)
\ee
with
\be
\label{3.3}
V (r) = - e  U_a \cdot \lk \frac{r}{r_a} \right)^2 = - \frac{m}{2} \omega^2_c r^2 \, .
\ee
We also define
\begin{align}
 \label{3.4}
 \omega_H &= \frac{eH}{2 mc} \, , \quad \quad \mbox{Larmor frequency} \, , \\
 \label{3.5}
 \omega_c &= \sqrt{\frac{2 e U_a}{m r^2_a}} \, , \quad \quad \mbox{critical frequency} \, .
\end{align}
Because
\begin{align}
 \frac{\partial H}{\partial t} &= 0 \, : \quad \quad H = E = const. = \alpha_1 \, , \nonumber\\
 \frac{\partial H}{\partial \varphi} &= 0 \, : \quad \quad p_\varphi = const. = \alpha_2 \, , \nonumber
\end{align}
 the Hamiltonian takes the form
\be
\label{3.6}
H = E = \frac{1}{2 m} \lk p^2_r + \frac{1}{r^2}  p^2_\varphi \right) + \omega_H p_\varphi + \frac{m}{2}
\lk \omega^2_H - \omega^2_c \right) r^2 \, ,
\ee
which, when solved for $p_r$, yields
\be
\label{3.7}
p_r = m \sqrt{\omega^2_H - \omega^2_c} \frac{1}{r} \sqrt{\frac{(2 m E - 2 m \omega_H p_\varphi) r^2}{m^2 (\omega^2_H - \omega^2_c)} - 
\frac{p^2_\varphi}{m^2 (\omega^2_H - \omega^2_c)} - r^4} \, .
\ee
Writing
\begin{align}
 a &= \frac{p^2_\varphi}{m^2 (\omega^2_H - \omega^2_c)} \, : \sqrt{a} = \frac{|p_\varphi|}{m \sqrt{(\omega^2_H - \omega^2_c)}} 
 \, , \quad \quad p_\varphi < 0 \, , \nonumber\\
 b &= \frac{(E - \omega_H p_\varphi)}{(\omega^2_H - \omega^2_c)} \nonumber
\end{align}
  we obtain for the action variable $J_r$:
  \be
  \label{3.8}
  J_r = \frac{m}{2} \sqrt{\omega^2_H - \omega^2_c} \frac{1}{2 \pi} 2 \pi (b - \sqrt{a})
  \ee
  or
  \be
  \label{3.9} 2 J_r = \frac{E}{\sqrt{\omega^2_H - \omega^2_c}} - \frac{\omega_H p_\varphi}{\sqrt{\omega^2_H - \omega^2_c}} - |p_\varphi| \, .
  \ee
 The action variables are defined as
 \be
 \label{3.10}
 J_\varphi = \frac{1}{2 \pi} \oint p_\varphi d \varphi \, , \quad \quad J_r = \frac{1}{2 \pi} \oint p_r d r \, .
 \ee
$J_r$ ist always positive. $J_\varphi$   positive or negative, according to the initial condition $p_\varphi \begin{array}{c} <0\\>0 \end{array}
$. For us, 
$J_\varphi = - |p_\varphi|, p_\varphi < 0$. 

Our result so far is given by
\be
\label{3.11}
H = E = 2 \sqrt{\omega^2_H - \omega^2_c} J_r + \lk \omega_H -\sqrt{\omega^2_H - \omega^2_c} \right) J_\varphi \, ,
\ee
which can be employed to define the two important frequencies
\begin{align}
 \label{3.12}
 \frac{\partial H}{\partial J_r} &= 2 \sqrt{\omega^2_H - \omega^2_c}  = \omega_r \, , \\
 \label{3.13}
 \frac{\partial H}{\partial J_\varphi} &= \omega_H -  \sqrt{\omega^2_H - \omega^2_c}  = \omega_\varphi \, .
\end{align}
The motion of the electron is now obtained in analogy to (\ref{2.13}) 
as a superposition of two uniform circular motions:
\be
\label{3.14}
r e^{i \varphi} = \sqrt{\frac{1}{m \sqrt{\omega^2_H - \omega^2_c}}}  \lk
\sqrt{(J_r - J_\varphi)} e^{i w_\varphi} - i \sqrt{J_r} e^{i (w_r + w_\varphi)} \right) \, .
\ee
This formula represents a particle in rolling motion on a circle with radius
\be
\label{3.15}
R_{rol} = \sqrt{\frac{1}{m \sqrt{\omega^2_H - \omega^2_c}}}  \sqrt{J_r} 
\ee
the center of which revolves on a (larger) circle with radius
\be
\label{3.16}
R_{rev} = \sqrt{\frac{1}{m \sqrt{\omega^2_H - \omega^2_c}}}  \sqrt{J_r -  J_\varphi} 
\ee
thereby generating an epicycloid, a picture that is well known from Ptolomaic astronomy and early atomic physics.

A similar calculation that produced equation (\ref{2.11}) is slightly modified and gives for the present case
\be
\label{3.17}
r^2 = \frac{2}{m \sqrt{\omega^2_H - \omega^2_c}} \left[\lk J_r - \frac{1}{2} J_\varphi \right) + \sqrt{J_r (J_r - J_\varphi)} 
\sin w_r \right] \, .
\ee
In equation (\ref{3.14}) and (\ref{3.17}) we have to add the relations
\begin{align}
 \label{3.18}
 w_r &= \omega_r t + \delta_r \, , \quad \quad w_\varphi = \omega_\varphi t + \delta_\varphi \, \\
 \label{3.19}
 \omega_r &= 2 \sqrt{\omega^2_H - \omega^2_c} \, , \quad \quad \omega_\varphi = \omega_H - \sqrt{\omega^2_H - \omega^2_c}  \, .
\end{align}
To simplify matters further, we introduce the canonical transformation
\begin{align}
 \begin{array}{lll}
  J_1 & = J_r - J_\varphi \, , \\
  J_2 & = J_r \, ,
   \end{array}
\mbox{so that} \quad J_\varphi = J_2 - J_1 \, .\nonumber
\end{align}
With the aid of the generating function
\be
F_2 (q_1, q_2; P_1, P_2) = F_2 (w_r, w_\varphi; J_1, J_2) = w_r J_2 + w_\varphi (J_2 - J_1)\nonumber
\ee
we obtain
\begin{align}
 \frac{\partial F_2}{\partial w_r} &= J_2 \, , \quad \quad 
  \frac{\partial F_2}{\partial w_\varphi} = J_\varphi = J_2 - J_1 = J_r - J_1 \, , \nonumber\\
  w_1 &= \frac{\partial F_2}{\partial J_1} = - w_\varphi \, , \quad \quad w_2 = \frac{\partial F_2}{\partial J_2} = w_r + w_\varphi \nonumber
  \end{align}
or
\be
\label{3.20}
w_r = w_1 + w_2 \, , \quad \quad w_\varphi = -   w_1 \, .
\ee
Therefore we can write for the Hamiltonian
\be
H = \omega_r J_r + \omega_\varphi J_\varphi = \omega_r J_2 + \omega_\varphi (J_2 - J_1)  = - 
\omega_\varphi J_1 + (\omega_r + \omega_\varphi) J_2 \nonumber
\ee
and so obtain for the frequencies
\begin{align}
 \label{3.21}
 \omega_r &=  \frac{\partial H}{\partial J_r} = 2 \sqrt{\omega^2_H - \omega^2_c} \, , \quad \quad \omega_1 = 
 \frac{\partial H}{\partial J_1} = - \omega_\varphi = 
 - \omega_H + \sqrt{\omega^2_H - \omega^2_c} =: - \Omega_2 \\
 \label{3.22}
  \omega_\varphi &=  \frac{\partial H}{\partial J_\varphi} = \omega_H -  \sqrt{\omega^2_H - \omega^2_c} \, , \quad \quad \omega_2 = 
 \frac{\partial H}{\partial J_2} = - \omega_r + \omega_\varphi = 
  \omega_H + \sqrt{\omega^2_H - \omega^2_c} =: \Omega_1 \, . 
\end{align}
Now formulae (\ref{3.14}) and (\ref{3.17}) take on the rather simple expressions - not unlike (\ref{2.24}) and (\ref{2.23}):
\begin{align}
 \label{3.23}
 r e^{i \varphi} &= \sqrt{\frac{1}{m \sqrt{\omega^2_H - \omega^2_c}}} \lk \sqrt{J_1} - i \sqrt{J_2} e^{i (w_1 + w_2)} \right)
 e^{- i w_1} \, ,\\
 \label{3.24}
 r^2 &=  \frac{1}{m \sqrt{\omega^2_H - \omega^2_c}} \left[ J_1 + J_2 + 2 \sqrt{J_1 J_2} \sin (w_1 + w_2) \right] \, ,
\end{align}
where
\begin{align}
 \label{3.25}
 w_1 &= \omega_1 t + \delta_1 = - \Omega_2 t + \delta_1 \, , \quad \quad \Omega_2 = \omega_H - \sqrt{\omega^2_H - \omega^2_c} \, , \nonumber\\
 w_2 &= = \omega_2 t + \delta_2 = \Omega_1 t + \delta_2 \, , \quad \quad \Omega_1 = \omega_H + \sqrt{\omega^2_H - \omega^2_c} \, ,
\end{align}
so that
\begin{align}
 \label{3.26}
 w_1 + w_2 &=  (\Omega_1 - \Omega_2)t + const. = 2 \sqrt{\omega^2_H - \omega^2_c} + const. \nonumber\\
 &= \omega_r t + const.
\end{align}
 Incidentally, we could continue our calculations using the proper frequencies $\Omega_1$ and $\Omega_2$ as done in ref. \cite{3}. 
 However, we would miss numerous details when in the following we consider the free electron motion given by the 
 Hamiltonian (\ref{3.6}) under the influence of an oscillating electric potential. Nevertheless, one should read the paper
\cite{3} when interested in a short overview of the problem under discussion. Also articles mentioned under \cite{4}
and \cite{5} are worth reading.
 
\begin{figure}
 \centerline{
 \includegraphics[width=.6\textwidth]{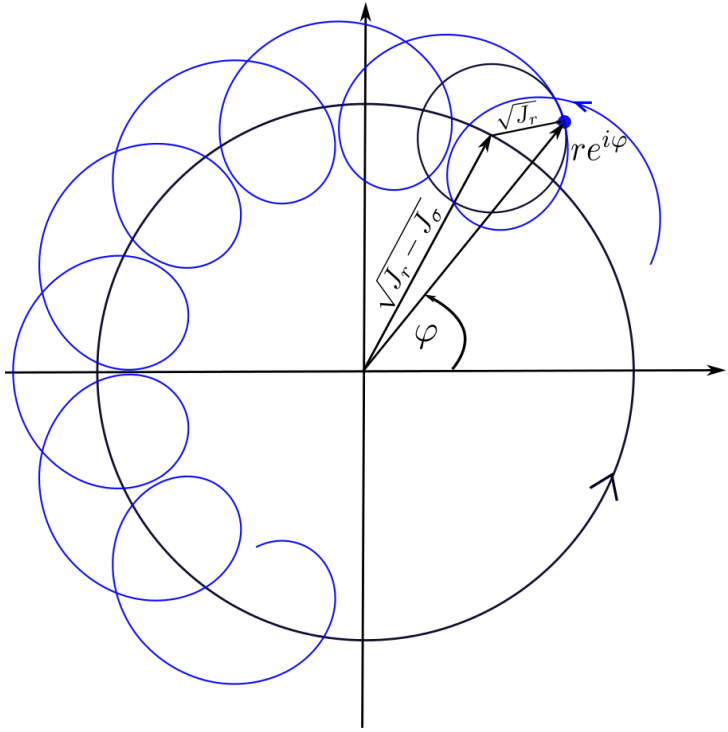}}
  \caption{Epicycloidic path of the electron.}\label{Fig.2}
\end{figure}

The two types of independent orbiting motion given in (\ref{3.15}) and (\ref{3.16}) can be written either way:
\begin{align}
 \label{3.27}
 R_{rol} &= \sqrt{\frac{1}{m \sqrt{\omega^2_H - \omega^2_c}}} 
 \sqrt{J_r} \quad or = \sqrt{\frac{1}{m \sqrt{\omega^2_H - \omega^2_c}}} 
 \sqrt{J_2} \, , \\
 \label{3.28}
  R_{rev} &= \sqrt{\frac{1}{m \sqrt{\omega^2_H - \omega^2_c}}} \sqrt{J_r - J_\varphi} 
  \quad or = \sqrt{\frac{1}{m \sqrt{\omega^2_H - \omega^2_c}}} 
 \sqrt{J_1} \, .
\end{align}
The distance of the electron from the center when it is most distant from the center is given by the aphelion radius
\be
\label{3.29}
R_{aph} = \sqrt{\frac{1}{m \sqrt{\omega^2_H - \omega^2_c}}} \lk \sqrt{J_r} + \sqrt{J_r - J_\varphi} \right) = 
\sqrt{\frac{1}{m \sqrt{\omega^2_H - \omega^2_c}}} \lk J_2 + J_1 \right) \, .
\ee
In fig. (\ref{Fig.2}) one gets a pictorial impression of the epicyclic motion of the electron. For an electron that starts initially
from the center we have to set
\be
\label{3.30}
J_{\varphi_0} = 0
\ee
and the energy is given by $E = J_r \omega_r$. The corresponding aphelion radius is given by
\be
\label{3.31}
R_0 = 2 \sqrt{\frac{1}{m \sqrt{\omega^2_H - \omega^2_c}}} \sqrt{J_r}  \, .
\ee

\section{Motion of the electron perturbed by an oscillating field}

So far we have considered the motion of an electron it its time development generated by the ``free'' 
Hamiltonian (\ref{3.6}). For this unperturbed motion we used the variables $J_r, J_\varphi; \delta_r, \delta_\varphi$, which are 
constant throughout the entire motion given by the unperturbed frequencies $\omega_r$  and   
$\omega_\varphi$. Now an external time-dependent oscillatory electric field is applied to the epicyclic motion.
  The electric potential of this field will be denoted by   $\phi (r, \varphi, t)$
  and can be expressed in the form of a Fourier expansion:
  \be
  \label{4.1}
  \phi (r, \varphi, t) = Re \sli^\infty_{\sigma = 1} \lk A_\sigma e^{- i \omega t} + B_\sigma e^{i \omega t} \right)
  \lk r e^{i \varphi} \right)^\sigma \, .
  \ee
Perturbing the electron by this potential will be of great importance when the electron orbit suffers a secular change 
at a resonant frequency. To find the condition for this to happen will be our next task.
When perturbed by the potential (\ref{4.1}), our former constant variables $(J_r, J_\varphi; \delta_r, \delta_\varphi)$ are 
no longer constant. Since these variables define a set of canonical coordinates and momenta, their 
time dependence is given by the canonical equations
\begin{align}
 \label{4.2}
 \dot{J}_r &= - \frac{\partial V}{\partial \delta_r} \, , \quad \quad \dot{J}_\varphi = 
 - \frac{\partial V}{\partial \delta_\varphi} \, , \nonumber\\
  \dot{\delta}_r &= \frac{\partial V}{\partial J_r} \, , \quad \quad \dot{\delta}_\varphi = \frac{\partial V}{\partial J_\varphi} \, .
  \end{align}
The potential energy $V$ of the electron is related to the alternating potential $\phi$ by
\be
\label{4.3}
V (J_r, J_\varphi, \delta_r, \delta_\varphi) = - e \phi (r, \varphi, t) \, .
\ee
As we know from section \ref{section2}, the coordinates $r$ and $\varphi$ can be expressed as functions of $(J_r, J_\varphi; w_r, w_\varphi)$. 
Since we are dealing with a time-dependent perturbation, we can Fourier transform both time and angle variables:
\begin{align}
 \label{4.4}
 V (J, w, t) &= Re \sli^\infty_{\sigma, \tau  = r} \Big[ C_{\sigma, \tau} (J_r, J_\varphi) \exp 
 \left\{ i (\sigma w_\varphi + \tau w_r - \omega) t \right\}  \nonumber\\
 &  + D_{\sigma, \tau} (J_r, J_\varphi) \exp \left\{ i (\sigma w_\varphi + \tau w_r + \omega) t \right\}  \Big]
\end{align}
or 
\begin{align}
 \label{4.5}
 V (J, \delta, t) &= Re \sli_{\sigma, \tau} \left[ C_{\sigma, \tau} \exp \left\{ i (\sigma 
 \delta_\varphi + \tau \delta_r) + i (\sigma \omega_\varphi + 
 \tau \omega_r - \omega)t \right\} \right. \nonumber\\
 & \left. + D_{c, \tau} \exp \left\{ i (\sigma \delta_\varphi + \tau \delta_r) + i (\sigma \omega_\varphi + 
 \tau \omega_r + \omega)t\right\} \right] \, .
\end{align}
Then the equations for the perturbed motion follow from (\ref{4.2}) and are given by
\begin{align}
\label{4.6}
 \dot{J}_\varphi &= - Re \sli_{r, t} \left[i \sigma C_{r, \tau} \exp \left\{
 i (\sigma \delta_\varphi + \tau \delta_r) + i (\sigma \omega_\varphi + i \omega_r - \omega) \tau \right\} \right. 
 \nonumber\\
 & \left. + i \sigma D_{c, \tau} \exp \{ \ldots + i \omega t \} \right] \nonumber\\
 \dot{J}_r &= - Re \sli_{\sigma, \tau} \left[ i \tau C_{\sigma, \tau} \exp \left\{ \ldots \right\} + i \tau D_{\sigma, \tau} \exp 
 \left\{ \ldots + i \omega t \right\} \right] 
 \nonumber\\
 \dot{\delta}_\varphi &= Re \sli_{\sigma, \tau} \left[ \frac{\partial  C_{\sigma, \tau}}{\partial J_\varphi} \exp \{ \quad \} 
 + \frac{\partial D_{\sigma, \tau}}{\partial J_\varphi}
 \exp \left\{ \ldots + i \omega t \right\} \right] 
\nonumber\\
\dot{\delta}_r &= Re \sli_{\sigma, \tau} \left[ \frac{\partial C_{\sigma, \tau}}{\partial J_r} \exp \{ \quad \} + 
 \frac{\partial D_{\sigma, \tau}}{\partial J_r} 
 \exp \{ \ldots + i \omega t \} \right] \, .
\end{align}
Because the right-hand side of (\ref{4.6}) are sums of periodic functions of $t$,
they give rise to small fluctuations of $(J_r, J_\varphi; \delta_r, \delta_\varphi)$, so that when averaged over a long time, 
they cause no substantial change of $J$ and $\delta$.
However, when the condition
\be
\label{4.7}
\sigma \omega_\varphi + \tau \omega_r - \omega = 0 \quad \quad \text{(resonance condition)} \, 
\ee
is fulfilled for some positive integral values of $\sigma$ and $\tau$, then some terms of the right-hand side 
in (\ref{4.6}) are no longer periodic and cause secular changes of $J$ and $\delta$. Consequently, the path of the 
electron deforms gradually and it becomes possible for the orbit of the electron, having started from 
the center and remaining initially in its neighborhood, to extend outwards until it reaches the anode.

Hence, after the averaging process, we obtain approximate equations of the secular changes of $(J_r, J_\varphi; \delta_r, \delta_\varphi)$, 
which are solution of
\begin{align}
 \label{4.8}
 \dot{J}_\varphi &= - Re i \sigma C_{\sigma, \tau}   (J_r, J_\varphi) \exp \left\{ i (\sigma \delta_\varphi + \tau \delta_r) \right\} \, , 
 \nonumber\\
  \dot{J}_r &= - Re i \tau C_{\sigma, \tau}   (J_r, J_\varphi) \exp \left\{ i (\sigma \delta_\varphi + \tau \delta_r) \right\} \, , 
  \nonumber\\
  \dot{\delta}_\varphi &= Re \frac{\partial C_{\sigma, \tau} (J_r, J_\varphi)}{\delta J_\varphi} 
  \exp \left\{ i (\sigma \delta_\varphi + \tau \delta_r)\right\} \, , 
  \nonumber\\
    \dot{\delta}_r &= Re \frac{\partial C_{\sigma, \tau} (J_r, J_\varphi)}{\delta J_r} 
    \exp \left\{ i (\sigma \delta_\varphi + \tau \delta_r) \right\} \, .
  \end{align}
In their simplified form, the above four equations now elucidate the secular change of the electron orbit. 
They clarify how the energy is conveyed from the electron to the oscillation and shed light on the process 
which forces electrons to move in phase.

To justify these remarks more specifically, let us study a four-split magnetron 
with oscillating voltages on four anode segments given by
\begin{align}
 \label{4.9}
 \phi (r_a, \varphi, t) = \cos \omega t \cdot \left\{
 \begin{array}{llll}
  \phi & , & \mbox{for} & 0 < \varphi < \frac{\pi}{2} \\
 - \phi & , & \mbox{for} & \frac{\pi}{2} < \varphi < {\pi}\\
  \phi & , & \mbox{for} & \pi < \varphi < \frac{3}{2} \pi \\
  - \phi & , & \mbox{for} & \frac{3}{2} \pi   < \varphi <  2{\pi}
     \end{array}
\right. \, .
\end{align}
We assume the gaps between the neighboring anode segments to 
be infinitely narrow. Also in the inner space the alternating potential is supposed to satisfy 
the the Laplace equation. This allows us to expand this potential in the form of a Fourier series:
\begin{align}
 \label{4.10}
 \phi (r, \varphi, t) &= \frac{4 \phi}{\pi} \cos \omega t \sli^\infty_{n = 0} \frac{1}{2 n + 1} \lk 
 \frac{r}{r_a} \right)^{2 (2 n + 1)} \sin 2 (2 n + 1) \varphi 
 \nonumber\\
 &= Re \frac{2 \phi}{\pi} \sli^\infty_{M = 0} \left\{ \frac{- i}{2 n + 1} \frac{1}{r^{2 (2 n + 1)}_a} r^{2 (2 n + 1)}
 \exp \left\{ i [2 (2 n + 1) \varphi - \omega t] \right\} \right. 
 \nonumber\\
  & + \frac{- i}{2 n + 1} \frac{1}{r^{2 (2 n - 1}_{a}} r^{2 (2 n + 1)} 
   \exp \left\{ i [ 2 (2 n + 1) \varphi + \omega t] \right\} \Big\}\, .
\end{align}
Notice that in this expansion only terms with $\sin 2 \varphi, \sin  6 \varphi, \sin 10 \varphi$, . . .
appear. Hence in the series (\ref{4.5}) and (\ref{4.6}) as well as in (\ref{4.8}), only terms with
\be
\label{4.11}
\sigma = 2, 6, 10, \ldots, 2 (2 n + 1), \ldots
\ee
appear. 

Using the equations (\ref{4.2}), (\ref{4.3}) and the Fourier series (\ref{4.10}), the coefficients $C_{\sigma, \tau}$
can  be calculated. 
For instance, for $\sigma = 2$, one finds
\begin{align}
 \label{4.12}
  C_{2, 0} &= \frac{2 e \phi}{\pi r^2_a} \frac{i}{m \sqrt{\omega^2_H - \omega^2_c}} (J_r - J_\varphi) \, , \nonumber\\
  C_{2, 1} &= \frac{4 e \phi}{\pi r^2_a} \frac{1}{m \sqrt{\omega^2_H - \omega^2_c}} \sqrt{J_r (J_r - J_\varphi)} \, , \\
  C_{2, 2} &= - \frac{2 e \phi}{\pi r^2_a} \frac{i}{m \sqrt{\omega^2_H - \omega^2_c}} J_r \, , \nonumber\\
  C_{2, \tau} &= 0 \quad \quad \mbox{for} \quad \tau \quad \mbox{other than} \quad \tau = 0, 1, 2 \, .\nonumber
\end{align}
In general it holds that
\begin{align}
\label{4.13}
C_{\sigma, \tau} & = 0 \quad \mbox{for values of} \hk \tau \hk  \mbox{other than} \nonumber\\
\tau &= 0, 1, 2, \ldots, \sigma \, .
\end{align}                                       
We know from the resonance condition (\ref{4.7}) that a secular change of the electron takes place when   
$\omega = \sigma \omega_\varphi + \tau  \omega_r$ is 
satisfied. Hence, $\sigma$  must take on one of the values given by (\ref{4.11}) and according to (\ref{4.13}) $\tau$ must be an 
integer number not greater than $\sigma$.

Now let us assume that the resonance condition holds true. Then for $\sigma = 2$ the allowed values for  
$\tau$ are given by $\tau = 0, 1, 2$.

For reasons given later we concentrate on $\sigma = 2, \tau  =0$. Then the equations given in (\ref{4.8}) yield
\begin{align}
 \label{4.14}
 \dot{J}_\varphi &= 2 \Omega (J_r - J_\varphi) \cos 2 \delta_\varphi \, , \quad \quad \dot{J}_r  = 0 \, , \nonumber\\
 \dot{\delta}_\varphi &= \Omega \sin 2 \delta_\varphi \, , \quad \quad \dot{\delta}_r = - \Omega \sin 2 \delta_\varphi \, , \nonumber\\
 \mbox{with} \quad \Omega &= \frac{2 e \phi}{\pi r^2_a} \frac{1}{m \sqrt{\omega^2_H - \omega^2_c}} \, .
\end{align}
These equations are easily integrated, leading to
\begin{align}
 \label{4.15}
 J_\varphi &= J_{\varphi_0} + (J_{\varphi_0} - J_{\varphi_{r_0}}) \left\{ \sin^2 \delta_{\varphi_0} e^{2 \Omega (t - t_0)} + 
 \cos^2 \delta_{\varphi_0} e^{- 2 \Omega (t - t_0)} \right\} \, , \nonumber\\
 J_r &= J_{r_0} \, , \nonumber\\
 \cos 2 \delta_\varphi &= - \frac{\sin^2 \delta_{\varphi_0} e^{2 \Omega (t - t_0)} - \cos^2 \delta_{\varphi_0}  
 e^{- 2 \Omega (t - t_0)}}{\sin^2 \delta_{\varphi_0} e^{2 \Omega (t - t_0)} + \cos^2 \delta_{\varphi_0} e^{- 2 \Omega (t - t_0)}} \, , \nonumber\\
 \cos 2 (\delta_r - \delta_{r_0} + \delta_{\varphi_0}) &= \cos 2 \delta_\varphi \, .
\end{align}
The quantities with index zero mean initial values. We stated already earlier under (\ref{3.30}) that 
as the electron starts initially from the center, we have to put
\be
\label{4.16}
J_{\varphi_0} = 0 \, .
\ee
With the results given in (\ref{4.15}) we can now see how the electron orbit deforms. 
To find this out, we merely have to substitute (\ref{4.15}) into (\ref{3.29}) to obtain
\be
\label{4.17}
R_{aph} = \frac{R_0}{2} \left[ 1 + \left\{ \sin^2 \delta_{\varphi_0} e^{2 \Omega (t - t_0)} + \cos^2 
\delta_{\varphi_0} e^{- 2 \Omega (t - t_0)} \right\} \right] 
\ee
with $R_0$  given by (\ref{3.31}).

Hence the aphelion radius increases until the electron reaches the anode. Note also that the ultimate increase 
of $R_{aph}$ is independent of the initial value of the phase of the electron motion. This emphasizes the fact
that all electrons will reach the anode independently of their initial conditions.
The electron energy can also be calculated. It is given by
\begin{align}
 \label{4.18}
 E &= E_0 + \omega_\varphi J_{r_0} \left[1 - \left\{ \sin^2 \delta_{\varphi_0} e^{2 \Omega (t - t_0)} + 
 \cos^2 \delta_{\varphi_0} e^{- 2 \Omega (t - t_0)} \right\} \right] \nonumber\\
 &= E_0 - 4 \omega_\varphi J_{r_0} \frac{R_{aph}}{R_0} \lk \frac{R_{aph}}{R_0} - 1 \right) \, .
\end{align}
With this equation, we can conclude that the energy of the electron decreases as it 
approaches the anode. This means that the net change of the electron while traveling from the cathode to the anode is given by
\be
\label{4.19}
\Delta E = - 4 \omega_\varphi J_{r_0} \frac{r_a}{R_0} \lk \frac{r_a}{R_0} - 1 \right) \, ,
\ee
which is clearly negative.
\bi

\no
This shows that during its travels, the electron is doing work. Due to the law of the conservation of energy, 
the electron passes this amount of energy on to the oscillation.

This leads us to the conclusion that if a tiny bit of oscillation happens to be generated in 
the oscillation circuit attached to the split anode, it will increase by the amount of energy delivered
by the electrons if the anode voltage and the applied magnetic field satisfy the resonance condition  
$\omega = 2 \omega_\varphi$. Since (\ref{4.19}) 
applies to all electrons, no matter what their initial condition, then all electrons add to the energy supply.
               
The third equation in (\ref{4.15}) yields a noteworthy finding: After a long enough time, $\cos 2 \delta_\varphi$
tends to $-1$, no matter what 
the initial value of $\delta_\varphi$ is. So while the values of the phases of the electrons' motions are initially quite random, 
after a while they become regulated and tend to  $\pi/2$ or $3 \pi/2$. If the initial value of $\delta_\varphi$
is in the first or second quadrant,
it tends to $\pi/2$; in the third or forth,  to $3 \pi/2$. 
As they approach the anode, the electrons gradually occupy the positions given by
\be
\label{4.20}
r e^{i \varphi} = r e^{i \lk \omega_\varphi t + \frac{\pi}{2} \right)} \quad \mbox{or} \quad r e^{i \lk \omega_\varphi
t + \frac{3}{2} \pi \right)} \, .
\ee
In this way they rotate with the angular velocity $\omega_\varphi = \omega_H - \sqrt{\omega^2_H - \omega^2_c} 
= \Omega_2$ as two clusters of space charge. The phase of rotation of the 
clusters causes positive charges in the anode segments with rising voltage and negative charges in those with 
decreasing voltage. So the rising voltage is increased even more and the decreasing voltage declines still further. 
Thus the rotating electron-cluster oscillations in the circuit intensify the oscillation in the circuit, and the 
nucleus of oscillation will expand until it reaches an intensity which can be detected. This self-excitation is
the fundamental principle behind how the magnetron functions.

Without any further calculations we want to mention that for the case  $\sigma  = 2, \tau =2$, the net change of energy is give by
\be
\label{4.21}
\Delta E = 4 (\omega_r + \omega_\varphi) \frac{r_a}{R_0} \lk \frac{r_a}{R_0} - 1 \right) \, ,
\ee
which is evidently positive, so that work is done by the alternating potential upon the electron; hence the 
oscillation in the circuit is weakened, i.e., when the resonance condition  $\omega  = 2 (\omega_\varphi + \omega_r) = 2 \lk 
\omega_H  + \sqrt{\omega^2_H + \omega^2_c} \right) = 2 \Omega_1$ is satisfied, no 
self-excitation of oscillation is possible.

For the remaining case  $\sigma = 2, \tau = 1$, it can be shown that the electron never reaches the anode, so
that no energy is supplied from the electrons to the oscillation. So again, no self-excitation of oscillation is possible.

These examples illustrate the secular change of electron orbits, 
how the electrons give or take energy from the oscillation and how electron spokes are created in the magnetron.
They show the three possible resonance scenarios: all electrons reach the anode by contributing energy to the oscillation; 
all electrons reach the anode by absorbing energy from the oscillation; or none of the electrons reach the anode, so 
there is no energy exchange.

This brings us to the close of our review of one of the most important inventions of the twentieth century. 
Unfortunately, there is not much textbook literature available on the theory of the magnetron. Of course, all 
the documents from the MIT Radiation Lab  that were classified during WWII have meanwhile been made public. 
But boundary-value problems in microwave physics are not easy to incorporate in an article that aims at explaining 
the theory and functioning of the magnetron.

Another remark should be made: Undoubtedly calculation of the Lamb shift or the unbelievably accurate computation 
of the anomalous magnetic moment of the electron by Schwinger and Tomonaga are highly impressive. Nevertheless,
it was a rather small group of physicists that hailed these results. But it was several million people who were 
influenced by the invention of the microwave generator for radar. We are all surrounded by super-high-frequency 
radio-wave devices, and the world would  look different without radar systems, be it in the physics lab, the airport or the kitchen.

One final comment might be of interest. After the detailed representation of the cavity magnetron that shows how 
electrons act on radio waves, I invite the reader to study the opposite, namely, how an external ``electrostatic'' 
plane wave would act on a charged particle. This is calculated in detail in chapter 11 of ref. \cite{1}, pages 147-156.

\end{document}